# Chondrules in enstatite chondrites

Emmanuel Jacquet[1], Laurette Piani[2], Michael K. Weisberg[3]

[1]Institut de Minéralogie, de Physique des Matériaux et de Cosmochimie, CNRS & Muséum National d'Histoire Naturelle, UMR 7590, 57 rue Cuvier, 75005 Paris, France.

[2]Department of Natural History Sciences, Hokkaido University, Sapporo 060-0810 Japan.

[3]Department of Physical Sciences, Kingsborough College, City University of New York, Brooklyn, NY 11235; Earth and Environmental Sciences, Graduate Center, CUNY, NY, NY 10016; Department of Earth and Planetary Sciences, American Museum of Natural History, NY, NY 10024. United States of America.

## Abstract
We review silicate chondrules and metal-sulfide nodules in unequilibrated enstatite chondrites (EH3 and EL3). Their unique mineralogical assemblage, with a wide diversity of opaque phases, nitrides, nearly FeO-free enstatite etc. testify to exceptionally reduced conditions. While those have long been ascribed to a condensation sequence at supersolar C/O ratios, with the oldhamite-rich nodules among the earliest condensates, evidence for relatively oxidized local precursors suggests that their peculiarities may have been acquired during the chondrule-forming process itself. Silicate phases may have been then sulfidized in an O-poor and S-rich environment; metal-sulfide nodules in EH3 chondrites could have originated in the silicate chondrules whereas those in EL3 may be impact products. The astrophysical setting (nebular or planetary) where such conditions were achieved, whether by depletion in water or enrichment in dry organics-silicate mixtures, is uncertain, but was most likely sited inside the snow line, consistent with the Earth-like oxygen isotopic signature of most EC silicates, with little data constraining its epoch yet.

## 1. Introduction

When thinking about the chondrule enigma, it is customary to concentrate on the common ferromagnesian, porphyritic chondrules found in ordinary or carbonaceous chondrites. Still, less usual chondrules (e.g. Al-rich, silica-rich, metallic…) can also offer their share of insight, for chondrule-forming scenarios, ideally, should account for them as well. There is a potential pitfall, however, in that nothing guarantees that all the objects we call "chondrules" formed by the same processes. For example, the currently widespread interpretation of the CB and CH chondrules as impact plume products (Humayun et al., this volume) does not necessarily hold for their more "mainstream" counterparts in other chondrite groups, which are very different petrographically and compositionally. Strangeness is an asset for the meteoriticist, but only within reasonable bounds.

Chondrules in enstatite chondrites (EC) may provide such a golden mean. Their mineralogy is exceptionally reduced, with normally lithophile elements locked in unique sulfide (e.g. oldhamite, niningerite, alabandite etc.) or nitride phases, high Si in metal, high abundances of opaque phases (including silicides, phosphides) overall, silica and FeO-free enstatite dominating over olivine. Yet the general textures and sizes of these chondrules are in the range seen in other chondrite clans (Jones

2012), suggesting relatively similar geneses. Their isotopic signature is closest to that of the Earth and the Moon than any other chondrite group, suggesting that they represent material from the inner Solar System and not some aberrant exotic reservoir. As further evidence that they are not mere curiosities, enstatite chondrites represent at least two independent chemical groups (EH and EL, with the former more metal-rich and reduced than the latter) plus several anomalous samples presumably from yet other parent bodies (Weisberg & Kimura 2012).

The reason for the unusually reduced character of enstatite chondrite chondrules is not known. It may actually pertain not only to silicate chondrules, but also to the opaque (metal-sulfide) nodules common in EC which it is a matter of language convenience to also call chondrules or not depending on their genetic links to the former. A difficulty in the study of unequilibrated EC (Weisberg & Kimura 2012) is their rarity, especially for observed falls (only three known, Qingzhen, Parsa and Galim, all EH3s), whereas finds are prone to the easy weathering of their reduced mineralogy. Yet substantial progress has been accomplished in the past decade on EC chondrules, from the isotopic, chemical and mineralogical point of view, with new insights on their astrophysical context of formation, and warrants a review of the field. In this chapter, we will describe the petrographic, chemical and isotopic characteristics of chondrules, including opaque nodules. We will then discuss competing models for their reduced parageneses, in particular formation in a reduced condensation sequence or reduction of material during chondrule melting. We finally discuss the astrophysical setting that may have brought about the fractionations inferred for such environments, and its possible spatio-temporal location in the early Solar System.

## 2. Petrography

### 2.1 Silicate chondrules

Chondrules make up ~60-80 vol% of enstatite chondrites (Scott & Krot 2014). Rubin & Grossman (1987) and Rubin (2000) determined an average chondrule diameter of about 220 μm for EH3 and 550 μm for EL3 chondrites. Schneider et al. (2002) reported similar sizes of 278 ± 229 μm and 476 ± 357 μm (average ± standard deviation) for EH3 and EL3, respectively. The largest chondrules seem to be more often nonporphyritic in texture (Rubin & Grossman 1987; Schneider et al. 2002). In the Sahara 97096 and Yamato 691 EH3 chondrites, Weisberg et al. (2011) found some chondrules up to 1 mm in diameter and one ~4 mm barred olivine chondrule. A broad correlation between chondrule and metal grain size in EH and EL chondrites may suggest aerodynamic sorting (Schneider et al. 1998). It is noteworthy that chondrules in E3 chondrites lack the fine-grained rims present in many other chondrites, and that compound chondrules are rare (Rubin 2010).

The chondrules in both EH3 and EL3 show a range of textures similar to those in ordinary and carbonaceous chondrites. However, enstatite (with generally <2 wt% FeO) is the major silicate phase in most E3 chondrules, with porphyritic pyroxene being the dominant chondrule texture (e.g., Fig. 1a, b). In general, the chondrules contain enstatite ± olivine (absent, or very minor, in many E3 chondrules) with a glassy mesostasis ± Ca-rich pyroxene ± minor silica (quartz, tridymite, cristobalite and glass; Kimura et al. 2005). Olivine is commonly, but not exclusively, poikilitically enclosed in enstatite (Fig. 1c). Although most of the chondrules in E3 chondrites may be characterized as type IB, they differ considerably from their counterparts in other chondrite groups, e.g. in the near pure endmember enstatite composition of their pyroxene or the occasional presence of silica.

Most strikingly, the EC chondrules contain a wide range of opaque phases unique to this clan (e.g. Grossman et al. 1985; El Goresy et al. 1988; Ikeda 1989b; Hsu 1998; Rubin 2010; Weisberg & Kimura 2012; Lehner et al. 2013; Piani et al. 2016). In EH3 porphyritic chondrules, Cr-Ti-bearing troilite, Si-bearing Fe-Ni metal, oldhamite (CaS) and niningerite ([Mg,Fe,Mn]S) are the most abundant opaque phases, with average modal abundances of 2.1, 0.7, 0.5, 0.4 vol%, respectively, for Sahara 97096 (Piani et al. 2016). Rare occurrences of minor caswellsilverite ($NaCrS_2$), two Cr-sulfides (minerals A and B of Ramdohr 1963), schreibersite ($[Fe,Ni]_3P$) and perryite ($[Fe,Ni]_8[Si,P]_3$) have also been reported (Grossman et al. 1985; El Goresy et al. 1988; Piani et al. 2016). Troilite occurs both enclosed in low-Ca pyroxenes and within the mesostasis associated with metal or other sulfides (Fig. 1d). Oldhamite and niningerite are often found together with troilite in micrometer-sized assemblages surrounded by mesostasis or connected to mesostasis channels between silicates (Piani et al. 2016). Niningerite is also frequently found in close association with low-Ca pyroxene, silica and troilite in silica-rich areas of porphyritic chondrules (Piani et al. 2016) and abundant niningerite and troilite were reported in silica-rich chondrules (Lehner et al. 2013). The EL3 chondrite chondrules also contain sulfides (Rubin 2010), typically oldhamite and alabandite ([Mn,Fe,Mg]S), with FeNi usually absent, but their textural and chemical properties and modal abundances have not been studied in detail.

Olivine (sometimes with evidence of reduction) and FeO-rich (up to 10.2 wt% FeO) pyroxene grains are present in some chondrules of the least equilibrated E3 chondrites (Rambaldi et al. 1983; Lusby et al. 1987; Weisberg et al. 1994), but are much less common than in ordinary or carbonaceous chondrite chondrules. No entire FeO-rich type II chondrules are present, but scarce ferroan pyroxene (up to 34 mol% ferrosilite) fragments (with silica[1]) and spherules have been described (Weisberg et al. 1994; Kimura et al. 2003; Jacquet et al. 2015). The rare olivine-rich porphyritic olivine/pyroxene (or olivine) chondrules in E3 chondrites (only 4 % of EC chondrules; Jones 2012) are similar to the type IA and IAB chondrules that are characteristic of ordinary and carbonaceous chondrites, but again may contain silica, Si-bearing metal and sulfides that are not found in the latter.

---

[1] Cristobalite and tridymite according to Raman spectra by E.J. on the Jacquet et al. (2015) objects.

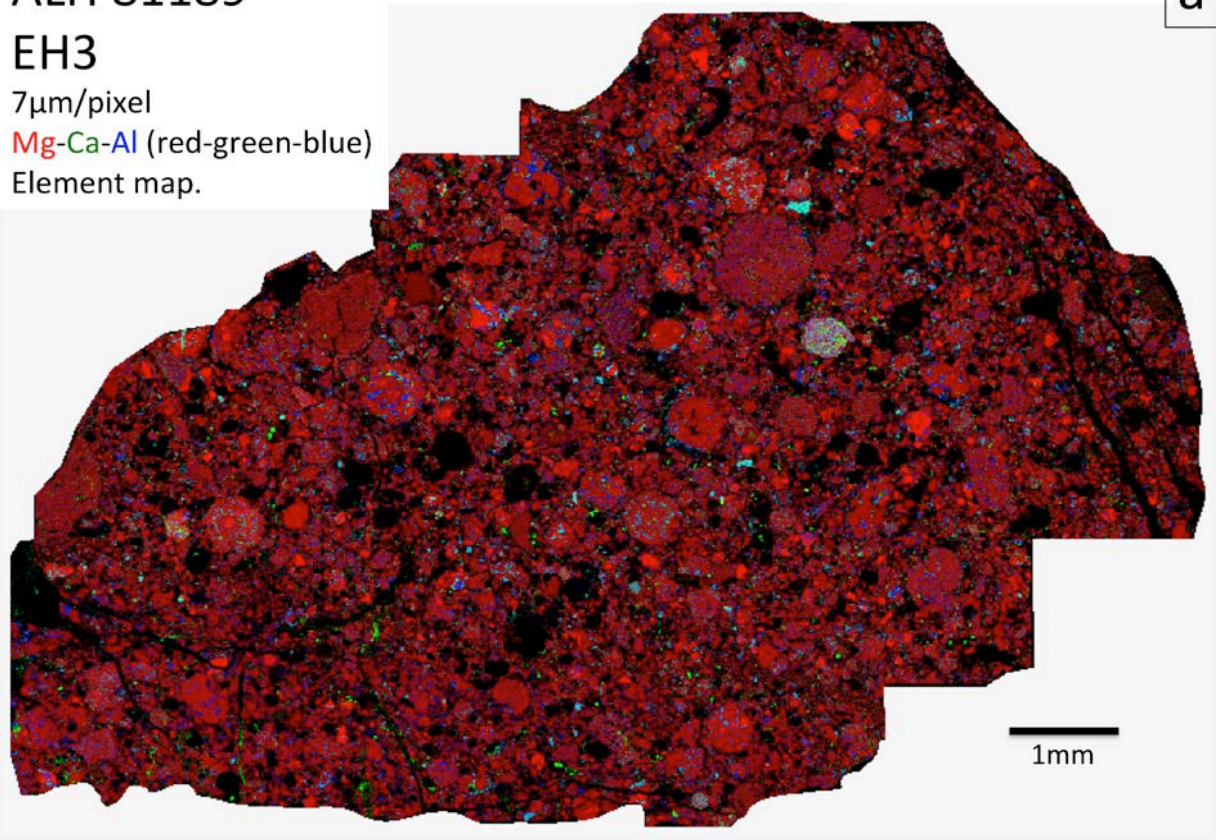

ALH 81189
EH3
7μm/pixel
Mg-Ca-Al (red-green-blue)
Element map.

1mm

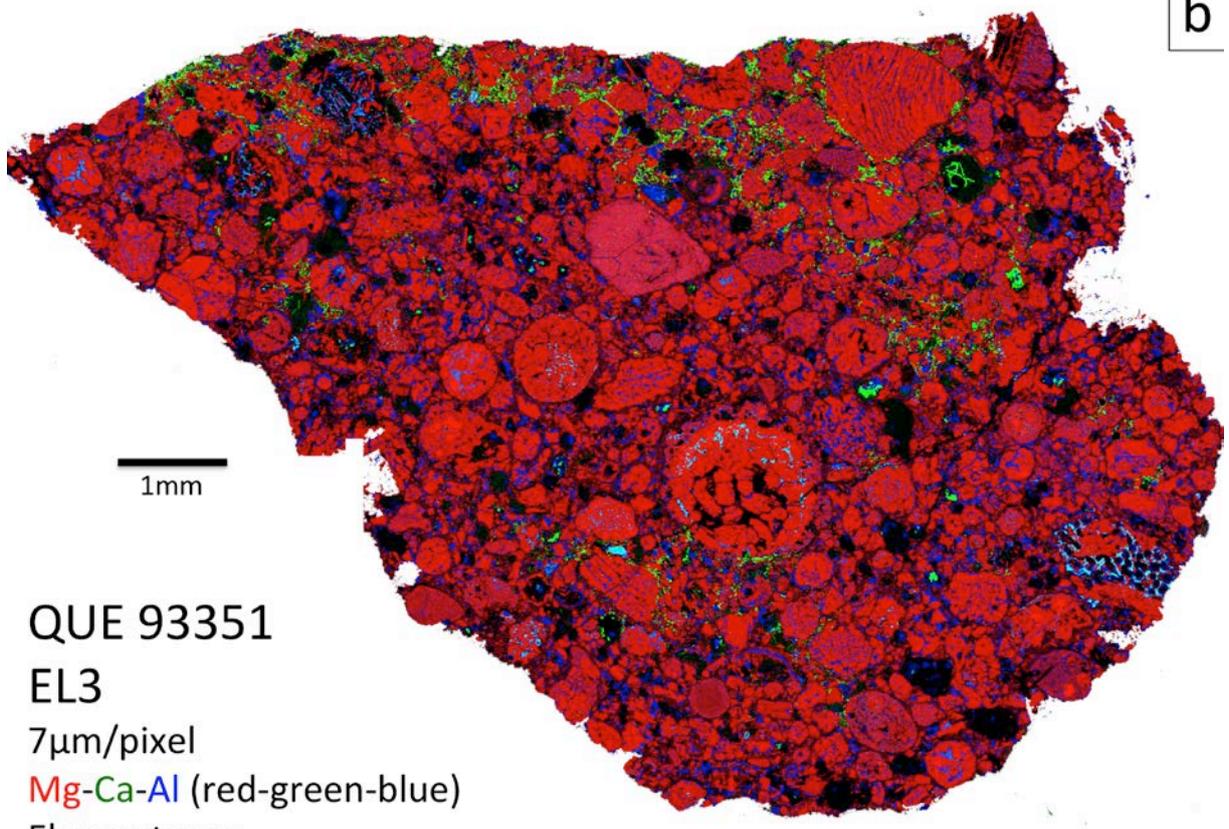

1mm

QUE 93351
EL3
7μm/pixel
Mg-Ca-Al (red-green-blue)
Element map.

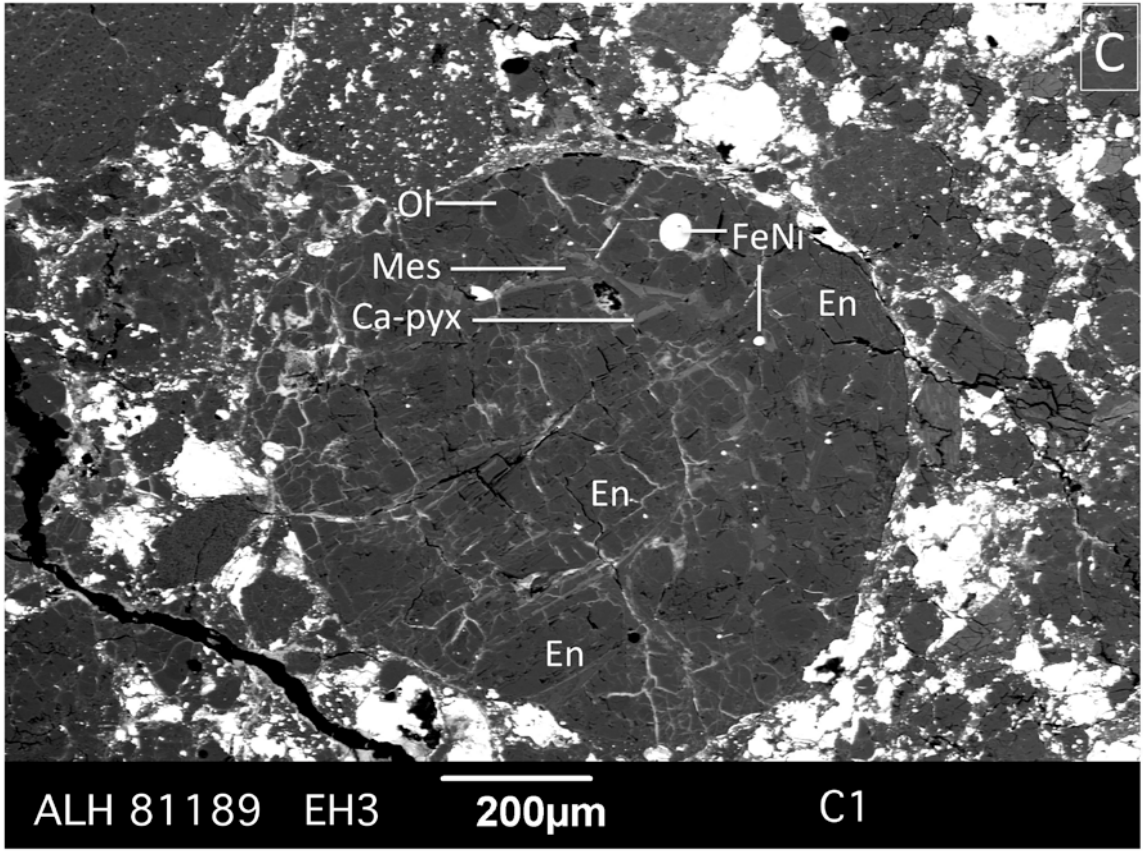

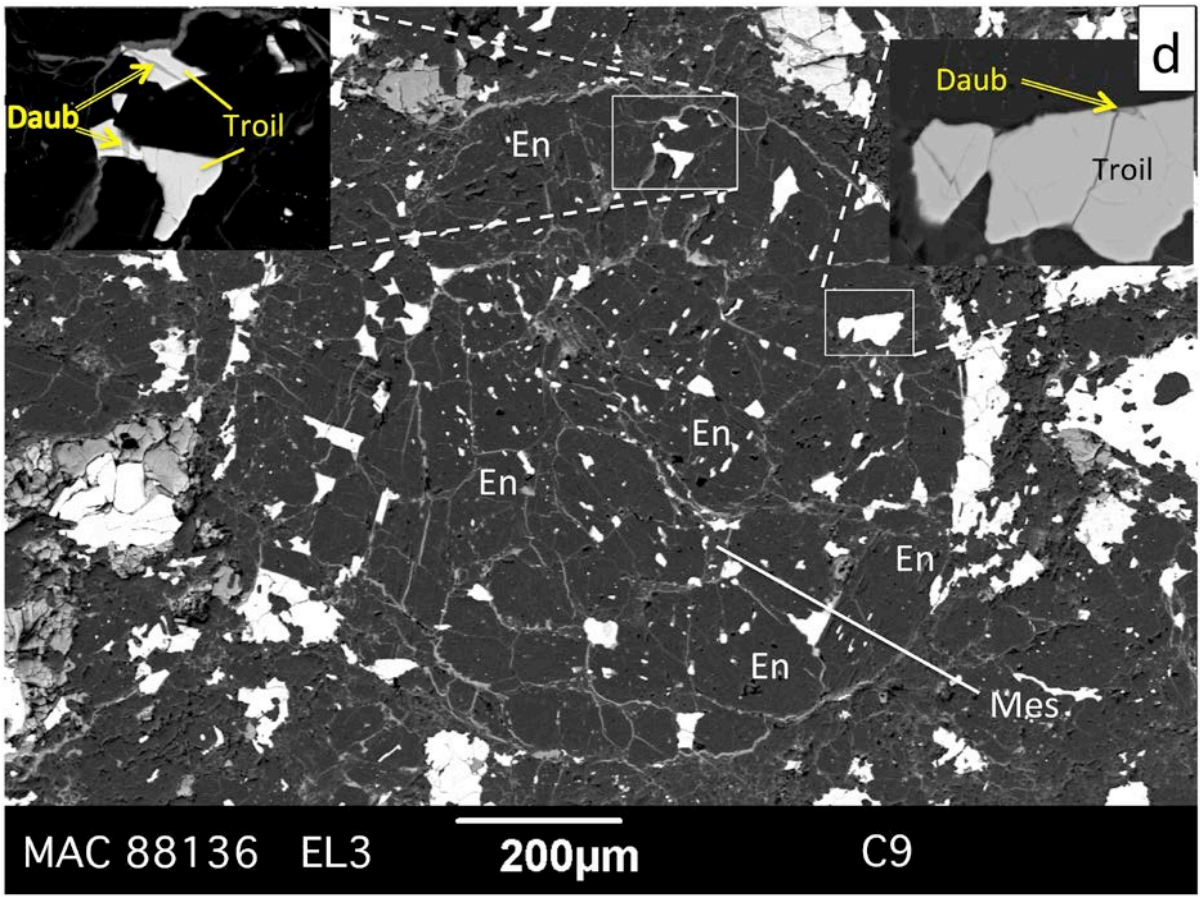

**Figure 1**: Mg-Ca-Al (red-green-blue) composite element map of the (a) ALH 81189 (,3) EH3 chondrite and (b) QUE 93351 (,6) EL3 chondrite showing chondrules of different textures and mineral abundances. In each image, the more common reds are dominantly enstatite, minor brightest reds are olivine and blues are feldspathic mesostases. Black regions are metal-sulfide nodules. (c) Example of a porphyritic chondrule from the ALH 81189 EH3 chondrite. The chondrule is dominantly enstatite (En) with poikilitic olivine (Ol), an albitic mesostasis (Mes) with Ca pyroxene (Ca-pyx) and minor Si-bearing FeNi. (d) Example of a porphyritic chondrule from the MAC 88136 (,37) EL3 chondrite dominated by enstatite ($Wo_{0.8}Fs_{0.5}$) with minor mesostasis and pyroxene-hosted troilite mixed with daubréelite (Troil + Daub) enlarged in the insets. Troilite has 0.36 wt% Cr and 0.29 wt% Ti.

## 2.2 Metal-sulfide nodules

In addition to silicate chondrules, the least equilibrated EL3 and EH3 chondrites contain abundant round to sub-round opaque assemblages ~50 to 800 µm in size (Weisberg & Kimura 2012; Lehner et al. 2010), often concentrically layered in EH3 chondrites (where they may make up 30-40 vol% of the meteorite; Weisberg & Prinz 1998; Lehner et al. 2014). Sometimes called "metal-sulfide chondrules" (e.g. Weisberg & Prinz 1998), they will be henceforth referred to as "metal-sulfide nodules" (MSN; Ikeda 1989a; Lehner et al. 2014).

Fe-Ni metal (kamacite with ~2-3 wt% of Si in EH3; <1.4 wt% in EL3) and troilite are the most abundant phases in MSNs in both EL3 and EH3 chondrites and are often associated with a variety of other sulfides, schreibersite, perryite, graphite and silicates.

In the EH3 chondrite MSNs, the other sulfides include oldhamite, niningerite, djerfisherite ($[K,Na]_6[Fe,Ni,Cu]_{25}S_{26}Cl$), caswellsilverite, daubréelite ($FeCr_2S_4$), and minor occurrences of other Cu and Cr-sulfides (e.g. Kimura 1988; Ikeda 1989a; Lin & El Goresy 2002). In the EL3 chondrite MSNs, these consist mostly of alabandite daubréelite, sphalerite and rare pendlandite ($([Fe,Ni]_9S_8$). Oldhamite was reported in inclusions in metal (Lin & El Goresy 2002) but may have been destroyed by terrestrial weathering in most EL3 chondrite samples.

Numerous silicates were reported in EH3 MSNs: low-Ca pyroxenes, silica (tridymite and cristobalite), roedderite ($[Na,K]_2[Mg,Fe]_5Si_{12}O_{30}$), albitic plagioclase (or albitic glass) and a porous amorphous silica (Keil 1968; Kimura 1988; Ikeda 1989a; Hsu 1998; Lehner et al. 2010). Silicates can be found as inclusions in kamacite, as intergrowths with sulfide and metal, or relatively continuous "garlands" (Fig. 2; Weisberg & Prinz 1998; Lehner et al. 2010). In most of the EL3 MSNs, euhedral to subhedral tabular or rod-shaped low-Ca pyroxenes form intergrowths with metal and sometimes sulfides (Weisberg et al. 1997; van Niekerk & Keil 2011; El Goresy et al. 2017, Fig. 2). Sinoite ($Si_2N_2O$) laths also sometimes occur associated with the low-Ca pyroxene (Horstmann et al. 2014, Fig. 2) and in sinoite-graphite-oldhamite assemblages in an EL3 chondrite clast in the Almahata Sitta polymict breccia (Lin et al. 2011).

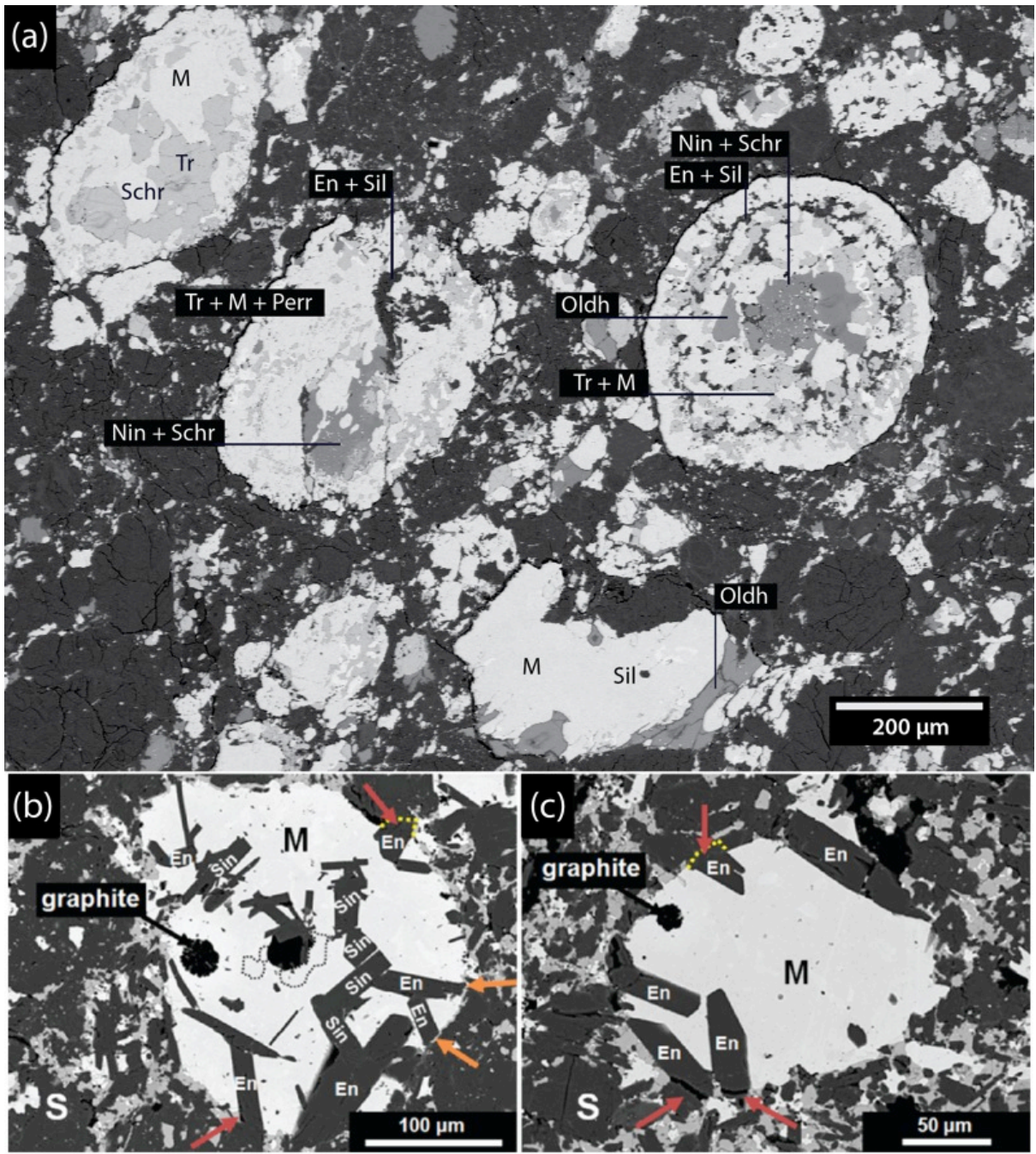

**Figure 2:** (a) MSNs in the EH3 chondrite Sahara 97096 and, (b) and (c) metal-silicate intergrowths with enstatite and sinoite laths, minor schreibersite exsolution (black dotted line in (b)) and graphite inclusions in the EL3/4 MS-17 from the Breccia Almahatta Sitta (both from Horstmann et al. (2014) to which the reader is referred to for details). M = Fe-Ni metal, Tr = troilite, Schr = schreibersite, En = enstatite, Sil = silica, Perr = perryite, Nin = Niningerite, Oldh = oldhamite, Sin = sinoite, S = silicates.

## 3. Chemistry

  Bulk silicate chondrule compositions in unequilibrated EC have been reported from analyses by instrumental neutron activation (Grossman et al. 1985), defocused/rastered electron microprobe (Ikeda 1988; Schneider et al. 2002), and laser ablation inductively coupled plasma mass spectrometry (Lehner et al. 2014; Varela et al. 2015). While the field of major element compositions mostly overlaps with chondrule populations from other chondrite clans (Jones 2012), some differences are apparent. Some of those pertaining to lithophile elements merely reflect the bulk meteorites (the matrix being a minor component): enrichment in Si (e.g. as normalized to Mg) and volatile elements like Na and Cl, impoverishment in refractory elements. A depletion relative to the whole rock in Ca, correlated with Eu and Se, seems to trace the loss of an oldhamite-rich component (Grossman et al. 1985), to which Varela et al. (2015) also ascribed negative (CI-normalized) anomalies in Nb, Ti, V, Mn in pyroxene-rich chondrules. EC chondrules are also uniformly low in siderophile elements (e.g. Fe, Co, Ni, Ir, Au) which are tightly correlated (Grossman et al. 1985) owing to their concentration in the opaque phases. Negative Eu, Yb and sometimes Sm anomalies are common (Lehner et al. 2014; see also leaching residue analyses by Barrat et al. (2014).

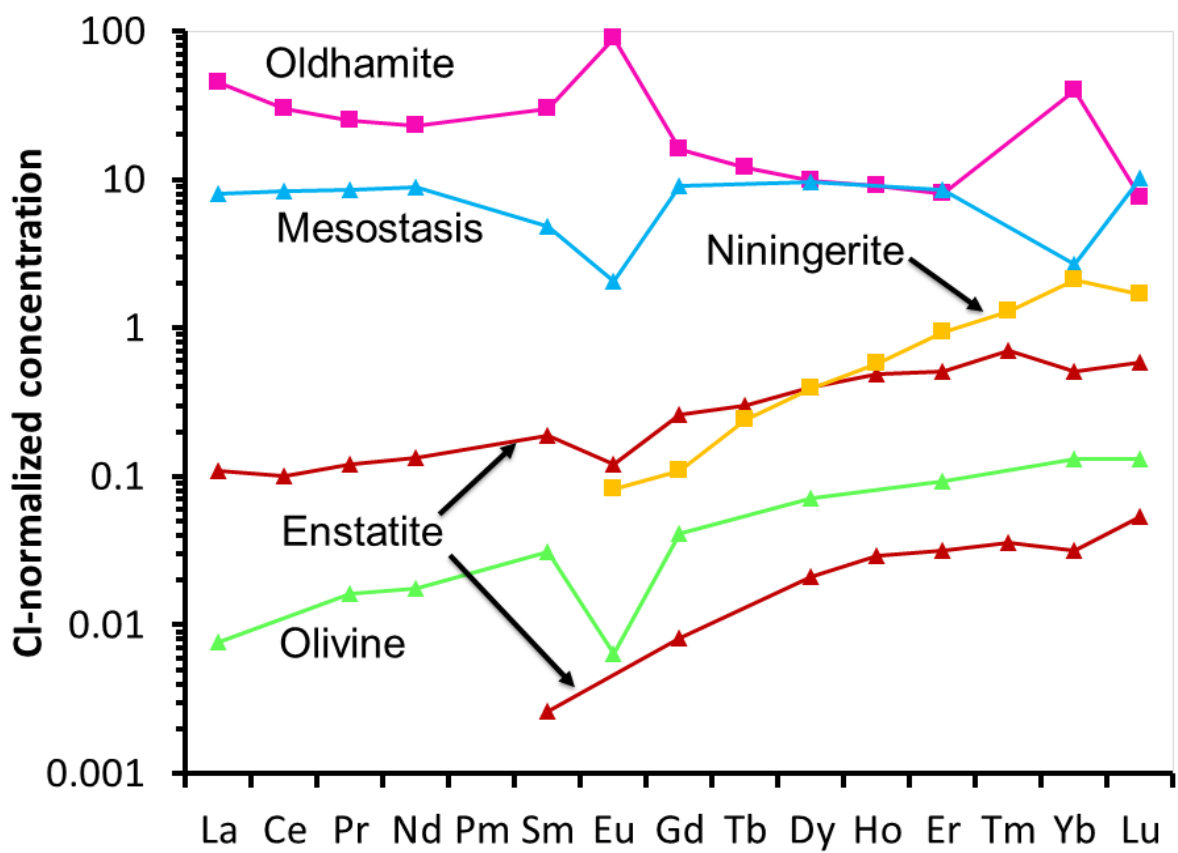

**Figure 3**: Representative rare earth element patterns of enstatite chondrite chondrule phases. Data sources are: Gannoun et al. (2011) for oldhamite (average of "type C-D" analyses in Sahara 97072); Crozaz & Lundberg (1995) for niningerite (average for Allan Hill 77156 (EH3/4)); Hsu & Crozaz (1998)

for enstatite (here their representative "pattern II" and "pattern III" for "blue enstatite"); Jacquet et al. (2015) for olivine and mesostasis (averages of Sahara 97096 chondrules).

If we turn to individual mineral compositions (reviewed by Brearley & Jones 1998), it is remarkable that minor element (Fe included) compositions of silicates (olivine, pyroxene) differ little, qualitatively, from those of type I chondrules in ordinary chondrites, although a population of enstatite with blue (instead of red) cathodoluminescence shows concentrations of all minor elements below 0.1 wt% (Brearley & Jones 1998; Schneider et al. 2002). This is true of their rare earth element (REE) patterns (Hsu & Crozaz 1998; Jacquet et al. 2015) as well, although negative Eu anomalies are more common (Fig. 3). In the most comprehensive (SIMS-based) study on pyroxene trace element chemistry in EC chondrules to date, Hsu & Crozaz (1998) found various levels of overabundance in light REE (relative to equilibrium pyroxene-melt partitioning expectations) which they ascribe to glass contamination. This may explain the flat patterns reported for enstatite by other workers (Weisberg et al. 1994; Crozaz & Lundberg 1995; Gannoun et al. 2011) and for ferroan pyroxene by Jacquet et al. (2015). Two porphyritic olivine-pyroxene (POP) chondrules as well as ferroan pyroxene spherules analyzed in Sahara 97096 exhibit *negative* REE pattern slopes which are as yet unexplained (Jacquet et al. 2015). The relative ordinariness of the silicates means that the bulk compositional peculiarities of EC chondrules are rather reflected in the modal mineralogy of the chondrules (e.g. the pyroxene/olivine ratio, sulfides) and the composition of the mesostasis (the main carrier of many incompatible elements). Compared to its ordinary chondrite counterparts, the latter is indeed depleted in Ca, Eu, Yb, Mn (Jacquet et al. 2015), and enriched in S (0.2-0.5 wt%; Piani et al. 2016), Cl (0-2 wt%; Grossman et al. 1985; Piani et al. 2016) and alkali like Na–more so in EH3 (5-13 wt% $Na_2O$; Grossman et al. 1985) than in EL3 (0-4 wt%; Schneider et al. 2002) although Manzari (2010) finds up to 9 wt% in Elephant Moraine 920299). Oldhamite, whether inside or outside silicate chondrules, is typically enriched by 1-2 orders of magnitude in REE relative to CI, with frequent positive anomalies in Eu and Yb in EH3 and negative Eu anomalies in EL3 chondrites (Larimer & Ganapathy 1987; Crozaz & Lundberg 1995; Gannoun et al. 2011; El Goresy et al. 2017; Fig. 3). Niningerite shows nearly chondritic heavy REE contents with monotonic depletion in light REE (Crozaz & Lundberg 1995; Gannoun et al. 2011; Fig. 3).

## 4. *Isotopic composition*

Oxygen isotope ratios of olivine and pyroxene in E3 chondrules show a wide range of values as large as 10 ‰ in both $\delta^{18}O$ and $\delta^{17}O$. Most chondrules plot along the terrestrial fractionation (TF) line on a 3-isotope diagram (Fig. 4a). However, some chondrules overlap the OC field and some extend toward more $^{16}O$-rich compositions. Tanaka and Nakamura (2017) interpreted the range of whole-chondrule O isotopic compositions they measured (including a silica-rich chondrule with a record-high $\delta^{18}O$ = 6.846 ‰) by varying mixtures between a carbonaceous chondrite-like precursor and nebular gas in the chondrule-forming environments. Most of the FeO-rich pyroxene in E3 chondrites plot on the TF line similar to enstatite, suggesting it formed locally in the EC region, indicating variable oxidation/reduction conditions within the E3 chondrule-forming region (Kimura et al. 2003; Weisberg et al. 2011). Olivine in some E3 chondrules shows a wide range of $\Delta^{17}O$ values (−4 ‰ to +2 ‰) and forms a distinct mixing line approximately parallel to but displaced from the carbonaceous chondrite anhydrous mixing (CCAM) line (Fig. 4a). Weisberg et al. (2011) referred to this as the "enstatite chondrite mixing line" but it is similar to the "primitive chondrule minerals line" defined by chondrules from the Acfer 094 chondrite (Ushikubo et al. 2012).

Weisberg et al. (2011) showed that some chondrules in EH3 chondrites have oxygen isotope heterogeneity among their minerals. These are interpreted to indicate incomplete melting of the chondrules, survival of minerals from previous generations of chondrules, and chondrule recycling. They found in particular one chondrule with a relict grain with a high, R chondrite-like $\Delta^{17}O$ value, suggesting limited admixtures of materials from other reservoirs.

From their oxygen isotope values, the chondrules in E chondrites seem to represent a distinct population formed in a local nebular reservoir different from chondrules in other groups but most closely related to ordinary and R chondrites. This is also shown by their $\Delta^{17}O$ values, which overlap with the range of compositions from LL3 and R3 chondrules. However, the majority of E3 chondrules have values close to 0 (i.e., Earth-Moon). In comparison, most LL3 and R3 chondrules have higher $\Delta^{17}O$ values while those in carbonaceous chondrites generally have negative $\Delta^{17}O$ values (Fig. 4b).

The isotopic composition of elements other than oxygen (e.g. Ca, Ti, Cr, Ni, Mo, Ru, W, Nd; see Burkhardt et al. (2017) and references therein) has been measured for bulk enstatite chondrites and generally found indistinguishable from the Earth. There are however few such data on *individual* chondrules. Gerber et al. (2017) found that the Ti isotope compositions of those chondrules were fairly uniform, around the terrestrial value, distinct from that of the (equally uniform) ordinary chondrite chondrules. This contrasts with the range exhibited in single carbonaceous chondrites, attributed by them to variable admixtures in their precursors of CAI-like material, presumably lacking in the enstatite chondrite reservoir.

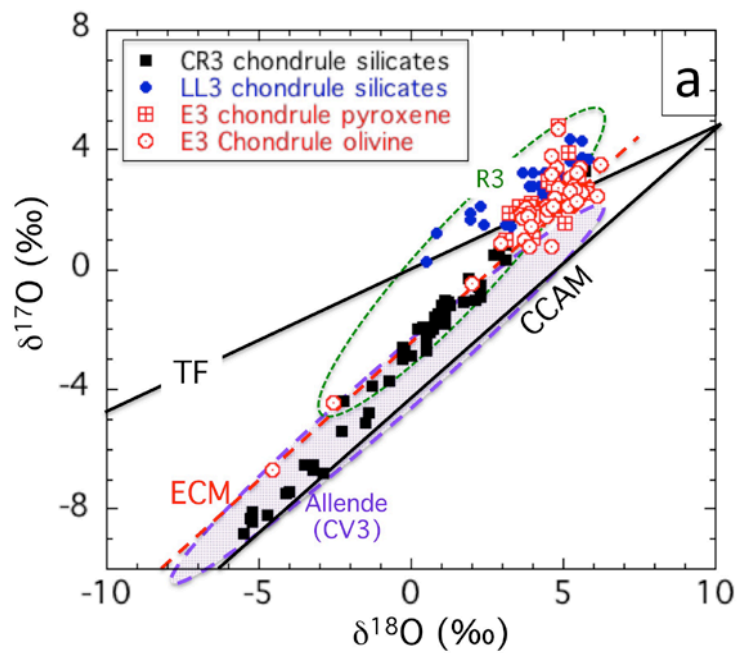

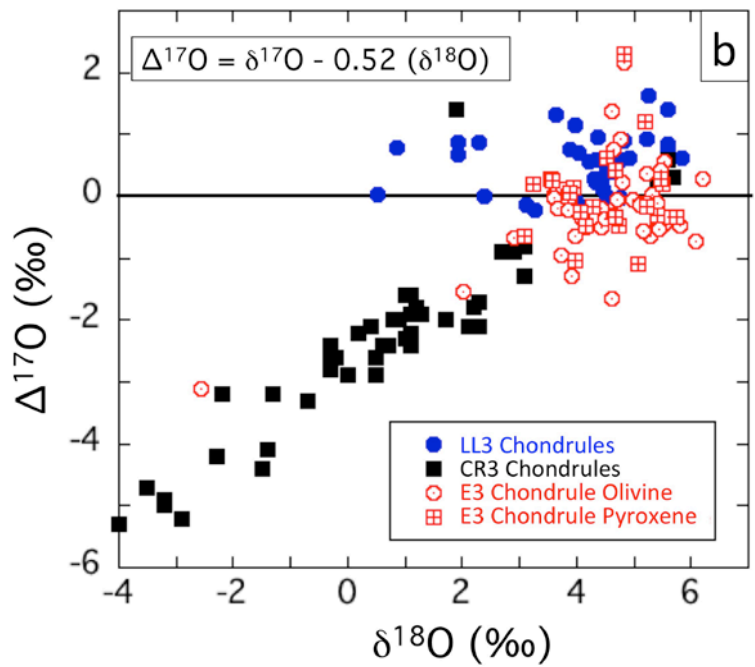

**Figure 4:** (a) Oxygen 3-isotope diagram showing SIMS data for chondrules in E3, LL3 and C3 chondrites. Also shown is the range of chondrule compositions in R3 chondrites and the Allende CV3 chondrite and the terrestrial fractionation (TF), carbonaceous chondrite anhydrous mineral (CCAM) and enstatite chondrite mixing (ECM) lines. Chondrules from E3 chondrites have compositions that overlap those from LL3 and R3 chondrites but their distribution appears to be unique with most data points clustered around the TF line. The E3 chondrules also fall along a line referred to as the ECM. The ECM is similar and possibly the same as the primitive chondrite mixing (PCM) line defined by chondrules in the primitive chondrite Acfer 094. (b) $\Delta^{17}O$ vs. $\delta^{18}O$ for chondrules in E3 chondrites compared to those of chondrules in LL3 and CR3 chondrites. Data are from Weisberg et al. (2011); Kita et al. (2010); Nakashima et al. (2010); Rudraswami et al. (2011); Tenner et al. (2013); Tenner et al. (2015).

## 5. A special condensation sequence…

   Clearly, the reduced mineralogy of enstatite chondrite chondrules (MSN included) sets them apart from other chondrite clans, but at which stage did their specificities essentially arise? Was it the formation of their precursors—presumably some nebular condensation sequence—or/and the chondrule melting process itself? Much of the past literature has envisioned the first option (Grossman et al. 2008; Lehner et al. 2013 and references therein). Some of it, in fact, envisioned chondrules in general, and those of EC in particular, as direct condensates (e.g. Blander et al. 2009; Varela et al. 2015), although support for this scenario has lately shrunk given the instability of liquids at the low total pressures expected in average protoplanetary disks (Ebel & Grossman 2000), the presence of relict grains and prevalence of porphyritic textures pointing to solid precursors (Hewins et al. 2005) or the few-Ma age spread of chondrules (Kita et al. 2013; Connelly et al. 2012), indicating repeated, short-lived formation episodes well beyond the "hot (inner) solar nebula" epoch. It is nonetheless also conceivable that chondrule-forming episodes essentially preserved the composition of the EC chondrule precursors, e.g. because of short timescales which would have averted devolatilization (especially in sulfur; Grossman et al. 1985) and for which Rubin (2010) finds supporting evidence in the small average diameter and lack of dust mantle of EC chondrules (in contradistinction to CV and CR type I chondrules, e.g.). The metal-sulfide nodules could also be nebular condensates (Rambaldi et al. 1986; Larimer & Ganapathy 1987; El Goresy et al. 1988; Kimura 1988; Ikeda 1989a; Lin & El Goresy 2002; Lehner et al. 2010), more or less affected by remelting and/or sulfidation events which may account for reverse zoning in some niningerite crystals (i.e. FeS increasing outward; Skinner & Luce 1971; El Goresy et al. 1988) .

Which condensation conditions would be required specifically? It has long been known that a condensation sequence of a solar mixture cannot account for the mineralogy of EC, in particular the dominance of pyroxene over olivine (even at low temperatures or pressures; Pignatale et al. 2016) or their array of sulfides–unless, perhaps, one assumes very high pressures (e.g. 1 bar) and iron supersaturation (Blander 1971; Herndon & Suess 1976; Blander et al. 2009). The stability of the reduced phases specific to EC is however possible if the C/O ratio is increased over the solar value of 0.5 (Larimer 1968; Baedecker & Wasson 1975; Larimer & Bartholomay 1979). This is because the stable species CO hereby locks an increasing fraction of the oxygen, hence lower $O_2$ fugacities. However, C/O ratios above ~1 would entail the condensation of phases such as carbides, graphite, nitrides which are but scarce in ECs (Larimer & Bartholomay 1979; Grossman et al. 2008); even though the then higher-temperature condensation of oldhamite would allow it to concentrate REE at the observed levels (Lodders & Fegley 1993), albeit with Eu and Yb anomalies of the *wrong* sign (negative) compared to observations (Gannoun et al. 2011). The relatively modest enhancements of Si contents in kamacite compared to carbonaceous chondrite metal also favor more moderate C/O ratios, perhaps around 0.8 (Grossman et al. 2008).

The subsolar Mg/Si ratios and refractory element abundances (e.g. Al/Si ratio) of EC chondrules (and whole-rocks), common to other non-carbonaceous chondrites, require fractionations among nonvolatile species too. Baedecker & Wasson (1975) and Hutson & Ruzicka (2000) postulated the loss of an olivine-rich refractory condensate. Petaev & Wood (1998) obtained this effect in a variant of fractional condensation calculation where 0.7-1.5 % of the solids equilibrating with the gas were isolated from it after each degree of cooling. Lehner et al. (2013) proposed Mg/Si fractionation of enstatite chondrites associated to the volatility of Mg in niningerite, which would however require another mechanism for the other non-carbonaceous chondrite clans (Jacquet et al. 2015).

## *6. … or special chondrule-forming conditions?*

Evidence is however mounting that EC chondrule precursors may actually have been less than remarkable. There is first no evidence for a particular CAI population: The rare CAIs in E3 chondrites are petrographically and isotopically similar to those in other chondrite groups; in particular, their oxygen isotopic compositions, even though statistically indistinguishable from the enstatite chondrite mixing line (Weisberg et al. 2011), are very $^{16}$O-rich and plot on the CCAM line, similar to CAIs in carbonaceous chondrites (Guan et al. 2000a; Guan et al. 2000b; Fagan et al. 2001; Lin et al. 2003). Second, the FeO-rich pyroxene and most relict olivine grains, with their fairly unexceptional chemistries (see "Chemistry" section above), share the same O isotope signature as the bulk EC (see "Oxygen isotopes" section); they cannot thus be mere interlopers from other reservoirs and more likely represent a population of EC chondrule precursors (or perhaps, an early stage of the chondrule-forming process itself). Third, Simon et al. (2016) found that 10-70 % of the Ti in E3 chondrules was tetravalent, although it is predicted to be largely trivalent for oxygen fugacities equal or lower than solar, indicating relatively oxidized precursors whose Ti valence was not efficiently reset.

As to metal-sulfide nodules, the patterns of siderophile elements measured in the metal of EH and EL MSNs do not correspond to those expected for a condensate but favor metal crystallization from a melt (Horstmann et al. 2014). Silicate laths intergrown with metal or sulfides in EL3 MSNs have also been interpreted as the results of melt crystallization, with van Niekerk & Keil (2011) suggesting impact on the parent body as the cause of the melting. Following Weisberg et al. (1997), Horstmann et al. (2014), noting the absence of textural indications for in situ melting (absence of melt pockets, FeS veins or intergrowths of the MSN phases within the matrix etc.) or significant REE redistribution between silicates and sulfides (Barrat et al. 2014), preferred impact prior to accretion. However, El Goresy et al. (2017) argue against an igneous origin of the MSN on the basis of their delicate mineralogical and C isotopic zoning or lack of quench textures. Still, euhedral pyroxene/metal intergrowths reminiscent of those are present in *EH* impact melt breccias like Abee and absent in EH3s (Rubin & Scott 1997), although their exact formation mechanism (as melted target material? or direct (liquid?) condensates from an impact plume? etc.) is unknown.

As to MSN in EH3 chondrites, the presence in oldhamite of numerous inclusions and intergrowths of silicate phases, often with euhedral shapes, also argues for an igneous origin (Hsu 1998). A further clue may be provided by their complementarity with silicate chondrules (Lehner et al. 2014; Ebel et al. 2015), in particular for siderophile elements, Ca (all depleted in the chondrules) and rare earth elements (with negative Eu and Yb anomalies; Fig. 3). The petrographic continuum between silicate-bearing metal-sulfide nodules to sulfide-bearing silicate chondrules (Lehner et al. 2014) also suggests a genetic link. Specifically, McCoy et al. (1999) suggested that the MSN were lost as immiscible droplets from the chondrules, in the same way opaques were likely expelled from molten chondrules in other groups, whether through surface tension or inertial acceleration effects (Grossman & Wasson 1985; Campbell et al. 2005; Jacquet et al. 2013), and may have further evolved by interaction with the gas. This has also been recently suggested for "sulfide chondrules" in unequilibrated R chondrites (Miller et al. 2017) and could be entertained for the MSN in LL3 chondrites (Weisberg et al. 2016) as well. Jacquet et al. (2015) proposed that igneous partitioning between oldhamite and the liquid chondrule mesostasis may explain the high REE content of oldhamite (as suggested for aubritic oldhamite by Dickinson & McCoy (1997)) as well as its frequent positive Eu and Yb anomalies if these elements were in divalent form. This would, of course, also account for the complementary negative Eu and Yb anomalies and Ca depletion of chondrule mesostases. Interestingly, a few ordinary chondrite chondrules present negative Eu, Yb and sometimes Sm anomalies (Pack et al. 2004; Metzler & Pack 2016; Ebert & Bischoff 2016) and are all CaO-poor (<0.8 wt%), suggesting that they, or at least their precursors, underwent similar fractionations. While Metzler & Pack (2016) objected that loss of oldhamite (light (L) REE-enriched) would have incurred unobserved LREE depletion in the chondrules, (i) the LREE enrichment of oldhamite is in fact moderate and likely further alleviated by accompanying phases such as (LREE-depleted) niningerite and (ii) the EC chondrule mesostases, in lacking the marginal LREE enrichment seen in other chondrite clans, do evidence some degree of relative LREE depletion (see Fig. 3). It is noteworthy that the EC chondrule mesostases' Eu and Yb anomalies are not reflected in the silicates, suggesting that the latter were already crystallized (whether formed at the beginning of the high-temperature episode in question or as relict of an earlier episode; Jacquet et al. 2015) when the opaque assemblages formed. Indeed, experimental heating in closed system of the Indarch EH4 chondrite indicates that the metal-sulfide components completely melt at 1000 °C (McCoy et al. 1999) and no sulfide (solid or molten) remains above 1400 °C in the presence of a silicate melt (Fogel et al. 1996; Fogel 1998), suggesting moderate temperatures at this stage.

What would the origin of sulfides in EC chondrules then be? The absence of niningerite and oldhamite enclosed in low-Ca pyroxene of the Sahara 97096 porphyritic chondrules indicates that these sulfides are not preserved from the chondrule precursors, unlike (possibly) olivine or Fe-Ni metal grains (Piani et al. 2016). At any rate, in order to avoid a rapid loss of sulfur from the molten chondrules (Yu et al. 1996; Uesugi et al. 2005), a high ambient partial pressure of sulfur is necessary (e.g. Marrocchi & Libourel 2013)—such as has long been suggested to explain the opaque paragenesis of enstatite chondrites (e.g. Fleet & MacRae 1987; El Goresy et al. 1988; Zhang et al. 1995; Lin & El Goresy 2002; Lehner et al. 2013). The stability of metal-sulfide mixtures, as a buffer of the $S_2$ fugacity, has been used by Lehner et al. (2013) to suggest S/H ratios 2,500-10,000 times higher than solar. Then, the sulfidation of silicates (olivine and pyroxene) by a S-rich gas at high temperatures (> 1000 °C) could result in the formation of Fe-, Ca-, or Mg-sulfides (Rubin 1983; Fleet & MacRae 1987; Lehner et al. 2013). Piani et al. (2016) noted that under the temperatures required for the sulfidation of solid silicates, with the chondrules being partially molten, the high partial pressure of sulfur should have interacted with the silicate melt. Thus sulfur would have dissolved in the silicate melt (hence the high S content of the mesostasis) and eventually saturated forming immiscible sulfide melts (e.g. Fincham & Richardson 1954; McCoy et al. 1999). The involvement of melt in the formation of niningerite may explain the large amounts of coexisting silica as well (Piani et al. 2016).

Another question is the origin of the high pyroxene/olivine ratio in EC chondrules compared to other chondrite clans. It may to some extent be a mechanical result of the bulk Mg/Si fractionation of the EC reservoir, whatever its origin (see end of previous section). It is however noteworthy that olivine in chondrules from other clans is predicted to dissolve within minutes in a melt of the composition measured in the mesostasis (Soulié et al. 2017). Libourel et al. (2006) suggested that the melt was enriched in Si by influx of SiO from the ambient gas, promoting pyroxene crystallization at the expense of olivine, and only a rapid cooling rate at that stage saved olivine (often rounded and partly resorbed) from wholesale dissolution. Then the lower olivine content in EC chondrules may perhaps reflect longer exposures to a SiO-rich gas and/or higher partial pressures of SiO compared to other chondrule-forming reservoirs. This has yet to be investigated.

In sum, it is conceivable that the specificities of EC chondrules (MSN included) were acquired from processing of relatively "normal" (chondrite-wise) precursors in unusually S- (and other volatile-)rich and O-poor environments. Then, the conditions of the condensation models discussed in the previous section may actually pertain to the chondrule-forming regions, for they would still adequately describe their equilibrium outcome by virtue of the path independence of equilibrium thermodynamics (notwithstanding lingering disequilibria). Whatever that may be, we have yet to find an astrophysical context for them.

## 7. Fractionation mechanisms

Although Grossman et al. (2008) entertain the possibility of primordial heterogeneities inherited from the presolar cloud, the weakness of the observed *isotopic* nucleosynthetic anomalies, even in refractory inclusions (Birck 2004), leaves little prospect of significant relative *chemical* anomalies at disk formation. One thus has to rely on dynamical processes within the protoplanetary disk to explain the nonsolar characteristics of the enstatite chondrite reservoir(s).

One way to increase the C/O ratio (or reduce $H_2O/H_2$) is a loss of rock and (if condensed) ice through gas-solid interaction. This could be effected e.g. by settling to the midplane (Ikeda 1989b; Krijt et al.

2016), or radial drift sunward (see simulations by Ciesla & Cuzzi 2006; Estrada et al. 2016). Another possibility would be efficient accretion of water and rock diffused beyond the snow line—a "cold finger"—, which, if more rapid than inward radial transport, could deplete the inner disk in water by orders of magnitude (Ciesla & Cuzzi 2006). Perhaps the S enrichment seen by enstatite chondrites would have been achieved beyond the troilite condensation front by similar "cold trapping" of troilite there as well (Pasek et al. 2005). Yet Grossman et al. (2008) object that these different oxygen-depleted regions would be even more depleted in the other condensable elements, and thus unsuitable to form EC components. Still, radial drift of early coarse refractory condensates would be a way to achieve the *nonvolatile* element fractionations of EC (Larimer & Wasson 1988; Jacquet et al. 2012).

A quite different scenario toward the wished reducing conditions would be a regional *enhancement*, by 2-3 orders of magnitude, in a *dry* mixture of rock and organic matter (some being still preserved in EC; see Piani et al. (2012) and references therein), e.g. an analog of "Chondritic Porous" (CP) interplanetary dust particles (Wood & Hashimoto 1993; Petaev & Wood 2001; Ebel & Alexander 2005). Indeed, even though the *global* C/O ratio may not necessarily be supersolar, the condensation of silicates would drastically deplete the *gas* in oxygen. This may also be a means to yield the sulfur enhancements proposed by Lehner et al. (2013). This requires a process which can efficiently fractionate water from the other condensable elements, most likely inside the snow line where the former is gaseous and the latter solid. Radial drift appears unable to produce enrichments above one order of magnitude in the inner disk (Ciesla & Cuzzi 2006; Estrada et al. 2016) and would likely include water from the outer disk (Jacquet & Robert 2013). One may thus be left with local processes such as turbulent concentration (Cuzzi et al. 2001) or vertical settling, the effect of which seems to be limited to one order of magnitude each in current models (Jacquet et al. 2012), although disk astrophysics admittedly remains quite uncertain.

One may then want to resort to non-nebular environments. The needed enrichments in condensable elements may come about in a planetary setting (Piani et al. 2016), e.g. an impact plume (Fedkin & Grossman 2013) on a reduced (chemically EC-like?) parent body (such as already inferred for metal-silicate intergrowths in EL3 chondrites; e.g. Horstmann et al. 2014). This, however, has yet to be investigated in detail from the modeling point of view in the specific context of enstatite chondrites.

## 8. Enstatite chondrite chondrules in space and time

However mysterious their context of formation remains, can we tell anything on where and when chondrules in enstatite chondrites formed? An origin inside the snow line, as suggested above, would be consistent with the lack of aqueous alteration in enstatite chondrites as wholes. Asteroids most spectroscopically consistent with enstatite chondrites, namely those of the E type, are most prevalent in the inner and middle parts of the main belt (DeMeo & Carry 2014). It is nonetheless unclear whether they lie significantly sunward of their S type counterparts, for the most conspicuous E asteroid population, among the Hungarias (inside 2 AU), may largely constitute a single Hirayama family (Milani et al. 2010), that is one initial parent body only. Yet, the putative mineralogy of Mercury is notoriously reminiscent of enstatite chondrites (McCoy & Nittler 2014), suggesting similar building blocks. Another hint at an inner Solar System provenance is the isotopic similarity in many elements of enstatite chondrites (and their chondrules) of both chemical groups to the Earth and the Moon. With the aubrites and the brachinite-like ungrouped achondrite Northwest Africa 5363, they may represent an "inner disk (isotopically) uniform reservoir" (Dauphas et al. 2014). Radial homogenization would indeed be more rapid at smaller heliocentric distances, given the shorter spatial scales and higher turbulent diffusivity.

The great degree of isotopic homogenization in the EC reservoir may also indicate a formation epoch later than other chondrite clans. This was also proposed by Jacquet et al. (2012) to allow for the Mg/Si and refractory element fractionation. Perhaps the relatively high proportion of type 3 in enstatite chondrites (especially EH) compared to ordinary chondrites (Scott & Krot 2014) may be traced back to then-weaker heat sources in their parent bodies at the time of their accretion. Direct radiochronological data on chondrules, while as yet consistent with a late melting, are however scarce: Individual I-Xe dating of EH3 chondrite chondrules yields ages of 4561-4564 Ma (Whitby et al. 2002); Guan et al. (2006) found evidence of *in situ* decay of $^{26}$Al in only one out of 13 Al-rich objects in unequilibrated enstatite chondrites but suggested disturbances by incipient metamorphism; and Trieloff et al. (2013) dated sphalerites in Sahara 97096 (EH3) by Mn-Cr at 4562.7±0.5 Ma.

## 9. Outlook

At the close of this tour of enstatite chondrite chondrule literature, we as a community are actually far from the end of the journey regarding the understanding of these objects. Still, with the progress accomplished in recent years, we may be able to ask better questions. So let us list some as concluding thoughts:

- Can we test further the possible link between silicate chondrules and the metal-sulfide nodules in enstatite chondrites?
- What are the petrographic characteristics of sulfides in EL3 chondrite chondrules and their relationships with silicates as compared to EH3?
- Under which redox conditions can experimental partition coefficients between oldhamite (and other sulfides) and silicate melts reproduce observations, especially as to the REE signatures? Is a condensate origin still possible for such signatures considering the observed overall mineralogy of enstatite chondrites?
- Why is olivine so rare in EC chondrules? Was it dissolved following Si influx in the chondrules during formation, and if so, what would the effect on their O isotopic composition be?
- Is it possible to increase the (water-free) condensable/gas ratio by 2-3 orders of magnitude above solar in the protoplanetary disk, as suggested by the S-rich composition of EC chondrules?
- Could such conditions be achieved in a planetary environment, e.g. an impact plume on a parent body with reduced mineralogy?
- Did EH and EL silicate chondrules and metal-sulfide nodules form by similar processes?
- How do EC chondrules fit in the absolute and relative chronology of chondrule formation in other clans?
- What are the parent bodies of EC chondrites? Where and when did they accrete?